\documentclass[pra,twocolumn,showpacs,nobibnotes,floatfix,superscriptaddress]{revtex4}

\usepackage{latexsym,eucal,amsmath,amssymb,amsfonts,mathbbol,graphicx,color}
\usepackage{dcolumn}
\usepackage{bm}
\usepackage{amsmath}

\begin{document}


\title{Pseudo-Random Circuits from Clifford Plus T-Gates}

\author{Yaakov S. Weinstein}
\affiliation{Quantum Information Science Group, {\sc Mitre},
200 Forrestal Rd., Princeton, NJ 08540, USA}

\begin{abstract}
We explore the implementation of pseudo-random single-qubit rotations and multi-qubit pseudo-random circuits constructed only from Clifford gates and the $T$-gate, a phase rotation of $\pi/4$. Such a gate set would be appropriate for computations performed in a fault tolerant setting. For single-qubit rotations the distribution of parameters found for unitaries constructed from Clifford plus $T$ quickly approaches that of random rotations and require significantly fewer gates than the construction of arbitrary single-qubit rotations. For Clifford plus $T$ pseudo-random circuits we find an exponential convergence to a random matrix element distribution and a Gaussian convergence to the higher order moments of the matrix element distribution. These convergence rates are insensitive to the number of qubits. 
\end{abstract}

     
\maketitle

Quantum information can be protected against errors by properly encoding it into suitable quantum error correction codes \cite{book}. Manipulating the information while it remains encoded can be done if all manipulations, such as quantum gates and the like, respect the symmetries of the code. The framework which will allows the implementation of a universal set of gates on the encoded information in such a way that the quantum information does not leave the encoded space is known as quantum fault tolerance (QFT) \cite{Preskill,ShorQFT,G,AGP}. Within a QFT setting many quantum error correction codes, such as the Calderbank-Shor-Steane (CSS) codes \cite{ShorQEC,CSS}, utilize a universal gate set consisting of Clifford gates, gates that map Pauli matrices to Pauli matrices, plus the $T$-gate, a single qubit $\pi/4$ phase rotation. Clifford gates can be implemented bit-wise, while the $T$-gate is implemented with the utilization of appropriate ancilla qubits. 

The universality of the gate set Clifford plus $T$, meaning the ability to implement any quantum operation using only gates from this set, does not by itself provide a prescription of how to use these gates to implement quantum protocols. A major difficulty in such a prescription is the implementation of arbitrary single-qubit rotations. Initial work on this problem was done in \cite{SK1,SK2} and more recently intense investigation has resulted in techniques with markedly improved efficiencies with respect to the number of necessary gates needed to achieve a prescribed gate accuracy $\epsilon$ \cite{Svore1,KMM1,TMH,Svore2,KMM2,Selinger1,KMM3}. In this paper we are interested not in implementing any specific gate, but in implementing random single qubit gates and random unitary operations with an arbitrary number of qubits with gates that are appropriate for QFT. Thus, it is necessary to design algorithms that can implement random unitary operators using only Clifford and $T$-gates. 

Random unitary operators and quantum states play an important role in many quantum information protocols. Random states saturate the classical communication capacity of a noisy quantum channel \cite{Seth2}, and are used for superdense coding of quantum states \cite{Aram}, and data hiding schemes \cite{Hayden}. Random quantum states can also be used for randomized benchmarking of quantum processes in the presence of noise \cite{MGE}. Random unitaries themselves are useful for remote state preparation \cite{Bennet} and noise characterization \cite{RM,EL,Levi}. 

The above protocols require random unitary operators drawn uniformly from the Haar measure of the circular unitary ensemble (CUE). However, the number of quantum gates necessary to implement CUE random unitaries on a quantum computer grows exponentially with the number of qubits \cite{RM,Zyc}. A possible substitute for CUE random matrices is the pseudo-random (PR) unitaries introduced in \cite{RM}. PR unitaries have statistical moments that approximate those of CUE matrices.
 
An efficient means of implementing PR unitaries is via PR circuits \cite{RM,ODP,Znidaric,MSB,BWV}. PR circuits consist of an iterated set of one and two-qubit gates having certain degrees of freedom which are chosen at random. As an example, each iteration of the standard PR circuit introduced in \cite{RM} consists of a random rotation on each single qubit followed by controlled-phase (CZ) gates between all nearest neighbors. The three Euler angles that determine the single qubit rotations serve as the degrees of freedom for the PR circuit. They are chosen randomly and independently for each rotation. As more iterations are applied (using different single-qubit gates for each qubit and at each time step) the statistical properties of the total unitary operator implemented compare more favorably to the statistical properties of random unitaries. 

Subsequent studies of PR circuits have focused on the convergence of such algorithms to different statistical properties of random unitaries \cite{ELL,YSW1}. Ref.~\cite{BV} specifically demonstrates that such circuits can efficiently implement unitaries whose statistical moments up to order $k$ approximate that of the Haar measure, within any prescribed accuracy $\epsilon$, for arbitrary $k$. Additional work has been done on the choice of two-qubit gates \cite{Znidaric}, the choice of single-qubit gates \cite{BWV}, some aspects of the topology of the qubits \cite{Znidaric,MSB,WBV}, and the ability of such unitaries to efficiently construct states of generic entanglement \cite{ODP}. PR circuits have also been formulated for cluster-state quantum computation \cite{BWV,PCP}. 

Unlike previous work, here we restrict our gate set to those appropriate when operating within a QFT framework. Thus, we do not assume the ability to perform arbitrary single qubit rotations but instead limit our gate set to those that will keep quantum information within the quantum error correction encoding. Thus, we will attempt to construct random rotations and PR unitaries utilizing only single qubit Clifford gates, the $T$-gate, and the CZ gate. 


We first explore the construction of random single-qubit rotations using only the Clifford gates Hadamard, $H$, and phase, $P$, and the $T$-gate given by:
\begin{eqnarray}
H =
\frac{1}{\sqrt{2}}\left( 
\begin{array}{cc}
1 & 1 \\
1 & -1 \\
\end{array}
\right) \;\;
P = 
\left( 
\begin{array}{cc}
1 & 0 \\
0 & i \\
\end{array}
\right) \;\;
T = 
\left( 
\begin{array}{cc}
1 & 0 \\
0 & e^{i\frac{\pi}{4}} \\
\end{array}
\right).
\end{eqnarray}
One possible construction protocol would be to randomly apply one of these three gates at every time step, $t$. However, we reject this suggestion as there are too many combinations of gates that would would be extraneous: $T^2 = P$ and $H^2 = \openone$. Instead we look to the gate sequences commonly found in prescriptions of arbitrary rotation using only gates from the set Clifford plus $T$ \cite{Svore1,Selinger1,KMM3}. We choose the sequences $HT$ and $PHT$ and apply one or the other at every time step to construct our single-qubit rotations. We equally weigh every one of the $2^t$ possible rotation for every time step up to $t = 25$ and compare the statistics of these unitaries with those of random single-qubit rotations. 

Random single-qubit rotations are unitaries drawn uniformly with respect to the Haar measure of $SU(2)$ and are completely parameterized by the Euler angles $\psi$, $\chi$, and $\phi$, as follows:
\begin{eqnarray}
U_1 &=& 
\left( 
\begin{array}{cc}
e^{i\psi}\cos\phi & e^{i\chi}\sin\phi \\
-e^{-i\chi}\sin\phi & e^{-i\psi}\cos\phi \\
\end{array}
\right),
\end{eqnarray}   
where $\psi$ and $\chi$ are drawn independently and uniformly from between 0 and $2\pi$, and $\phi = \sin^{-1}\sqrt{\xi}$ where $\xi$ is drawn uniformly from between 0 and 1. 

To compare the Clifford plus $T$-gate constructed unitaries with single-qubit random unitaries we extract from each of the $2^t$ constructed unitaries the paramaters $\psi$, $\chi$, and $\xi$ which are then sorted into equally spaced bins (our simulations are only slightly dependent on the number of bins). The normalized distributions of these parameters, $\tilde{P}(\alpha)$ for $\alpha = \psi, \chi, \xi$, are compared to the appropriate distributions for random unitaries, $P(\alpha)$. The difference between these distributions is then calculated as $D(\alpha) = \sum|\tilde{P}(\alpha)-P(\alpha)|^2$ where the sum is taken over all bins. 

\begin{figure}[t]
\includegraphics[width=8cm]{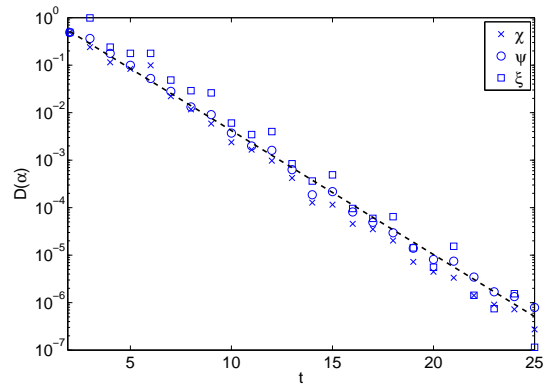}
\caption{Difference, $D(\alpha)$, between the distributions of $\xi$ ($\times$), $\psi$ ($\circ$), and $\chi$ ($\square$) for random single-qubit unitary matrices and those constructed from Clifford plus $T$-gates as a function of time step, $t$. The least squares fit to $D(\psi)$ (dotted line) is given by $\exp(.54-.60t)$. Least square fitting to $D(\chi)$ and $D(\xi)$ give similar coefficients. 
}
\label{Q1}
\end{figure}

Fig. \ref{Q1} plots each $D(\alpha)$ as a function of time step. As shown, $D(\alpha)$ decreases at an exponential rate $e^{-\kappa t}$ where a least squares fit for the decay constant $\kappa$ gives .63, .60, and .66 for $D(\chi)$, $D(\psi)$, and $D(\xi)$ respectively. These results demonstrate that the distribution of single qubit rotations based on Clifford plus $T$ gates quickly approaches that of random unitaries justifying our initial choice of gate to sequences to be applied. We note that the average number of time steps to achieve $D(\alpha) < 10^{-5}$ is 20 which translates into 20 $T$-gates and an average total of 50 single qubit gates. This number is significantly below the number of single qubit gates needed to construct an arbitrary single qubit rotation to the same accuracy \cite{Selinger1,KMM3}. 

Based on the above, a straightforward way to implement PR circuits on multiple qubits using only gates from Clifford plus $T$, is to simply replace the single-qubit unitaries drawn from $SU(2)$ of the standard PR circuit \cite{RM} with a sequence of $HT$ and $PHT$ gates that would implement a PR single-qubit unitary. The convergence to CUE statistics would be similar to the standard case at an increased cost in number of gates applied equal to the number of gates used to implement the PR single-qubit rotation (depending on the desired accuracy) times the number of qubits. 

For the sake of increased efficiency, however, we would like to explore the possibility of applying only one iteration of the sequence $HT$ or $PHT$ on each of the qubits in place of the random single-qubit rotations of the standard PR circuit. Thus, a time step, $t$, of the Clifford plus $T$-gate PR circuit on a line of $n$ qubits would involve applying to each qubit either the single qubit gates $HT$ or the gates $PHT$ (each with a probability of .5), followed by CZ gates between all nearest neighbors. 

To determine the randomness of the PR unitaries constructed in this way we look at the matrix element distribution and its higher order moments (as in \cite{AB}). For CUE matrices, random matrix theory provides the following distribution:
\begin{equation}
P(l) = \frac{N-1}{N}e^l\left(1-\frac{e^l}{N}\right)^{N-2}
\end{equation} 
where $N = 2^n$ is the Hilbert space dimension and $l$ is a function of the matrix elements $U_{ij}$ given by $l = \ln(N|U_{ij}|^2)$. We compare this distribution to that of the PR unitaries from Clifford plus $T$-gates by constructing a sample number $r$ of PR unitaries and binning the $rN^2$ $l$ values into equally spaced bins. The distance between the normalized distributions is, as above, given by:
\begin{equation} 
D(l) = \sum|\tilde{P}(l)-P(l)|^2, 
\end{equation}
where the sum is taken over all bins. This is done for $n = 6, 8, 10, 12$ and 14 qubits using $r = 10000$ for the cases $n = 6$ and 8, $r = 1000$ for $n = 10$, and $r = 50$ for $n = 12$ and $r = 5$ for $n = 14$. 

\begin{figure}[t]
\includegraphics[width=8.5cm]{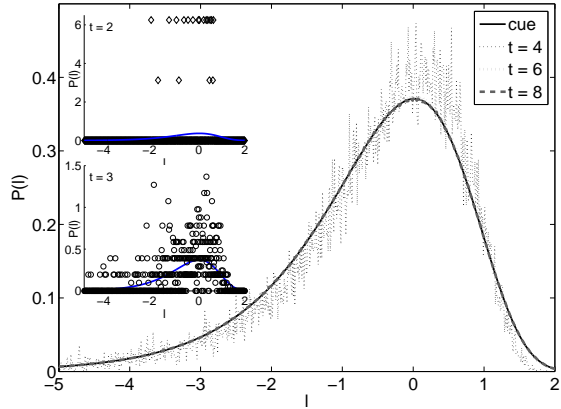}
\caption{Matrix element distribution, $\tilde{P}(l)$, for $n = 6$ PR circuits from Clifford plus $T$ gates at different time steps. For $t = 2$ (top left inset) and $t = 3$ (bottom left inset) the distribution is simply a series of large spikes. For higher $t$ (main figure) the spikes merge into a continuous distribution and collapse into the CUE random matrix element distribution.   
}
\label{App}
\end{figure}

The convergence of the matrix element distribution for the Clifford plus $T$-gate constructed unitaries to that of CUE 
is shown in Fig.~\ref{App} for the case of $n = 6$. Of note is the behavior of the approach. Initially the matrix elements are confined to very specific magnitudes such that the distribution is simply a series of large spikes. As $t$ increases the spikes shrink and increase in number before joining together to collapse into the desired distribution. This behavior should be contrasted, for example, with that demonstrated in \cite{AB} where for low $t$ the distribution is heavily weighted towards higher magnitude elements before spreading out and filling up the lower magnitude parts of the distribution. 

\begin{figure}[t]
\includegraphics[width=8cm]{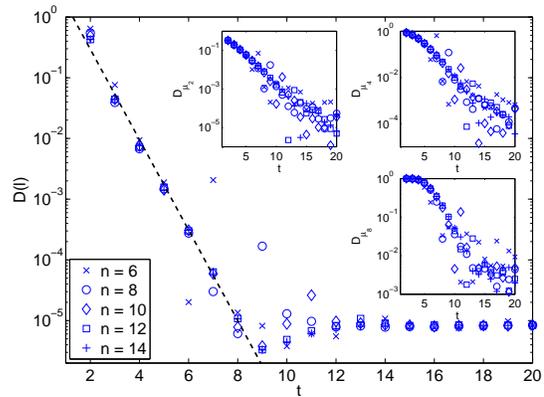}
\caption{Difference $D(l)$ between the distributions for CUE random matrices and those constructed via PR circuits from Clifford plus $T$ gates for $n = 6$ ($\times$), 8 ($\circ$), 10 ($\diamond$), 12 ($\square$), and 14 ($+$) qubits. The least squares fit to the 12 qubit case (dotted line) is given by $\exp(2.21-1.71t)$. The insets show the deviation from the random matrix derived moments of the matrix element distribution as a function of time step for moments $k = 2$ (top left), 4 (top right), and 8 (bottom). Note that the convergence to the moments is not exponential but Gaussian.  
}
\label{Q2}
\end{figure}

The complete results are shown in Fig.~\ref{Q2} and demonstrate the ability to construct PR unitaries from the Clifford plus $T$ gates. As the number of time steps increase $\tilde{P}(l)$ converges to $P(l)$ at an exponential rate marred only by an overshoot at $t = n$ followed by a spike at $t = n+1$. The magnitude of this overshoot and spike decreases with increasing number of qubits. In addition, the rate of convergence is independent of the number of qubits and the decay constant is $\kappa \simeq 1.71$ (this will depend somewhat on the level of binning). We note that the lack of sensitivity to qubit number is due to the PR circuit prescribing that two qubit gates are applied between all nearest neighbors at every time step. PR circuits that apply only one two-qubit gate at a time, such as \cite{AB}, converge at a rate that is strongly dependent on the number of qubits. To explore further the accuracy with which the PR unitaries built from Clifford plus $T$ gates resemble random unitaries we look at higher order moments of the matrix element distributions $\tilde{P}(l)$. 

Moments of distribution of matrix elements were analyzed in the context of PR unitaries in Ref.~\cite{AB}. Here we are especially intersted as to whether the evolution of these moments will depend on the number of qubits. The $k$th moment of the matrix element distribution $\mu_k$ is defined as $N^k\langle|U_{i,j}|^{2k}\rangle$. For CUE matrices the moments are given:
\begin{equation}
\mu_k = \frac{k!N^k(N-1)!}{(N+k-1)!}.
\end{equation}
We look at the deviation from the random matrix derived moments via 
\begin{equation}
D_{\mu_k} = \frac{|\mu_k - \tilde{\mu}_k|}{\mu_k},
\end{equation}
where $\tilde{\mu_k}$ is the calculated matrix element distribution moment for the unitaries constructed from the set of gates Clifford plus $T$. The results for moments $k = 2, 4$ and 8 are shown in the insets of Fig.~\ref{Q2}. First, we see that, as with the difference in distributions, the results are basically independent of the number of qubits except for the same overshoot and recovery phenomenon discussed above at time steps $t = n$ and $n + 1$. In contrast to the difference in distributions however, the rate of convergence to the CUE is not exponential but a Gaussian with the exact behavior depending on $k$: the higher the moment the slower the initial convergence.  

In conclusion, we have demonstrated the construction of random single-qubit unitaries by stringing together sequences of the gates $HT$ and $PHT$. We have shown that the statistical distributions of the Euler angles from the set of unitaries quickly approach that of random $SU(2)$ matrices. We then extended the exploration to more qubits, devising pseudo-random circuits utilizing only Clifford gates and the $T$ gate. The matrix elements from the unitaries thus constructed quickly approach the distribution of CUE matrix elements with little sensitivity to the number of qubits. 

This exploration provides a useful algorithm to construct random states and unitaries within the quantum fault tolerant framework. Future work will focus on the accuracy and robustness of the algorithm when subject to errors. In that case (noisy) gates will be implemented on logical qubits that allow quantum error correction to be explicitly implemented.    

YSW thanks S. Pappas for constructive conversations and G. Gilbert for comments. Support provided by the MITRE Technology Program under MTP grant \#07MSR205.

\end{document}